\documentstyle[12pt]{article}
\begin{document}
\begin{center}
{\Large\bf A-Dependencies of Neutral Strange Particle Yields
at $40~GeV/c \bar p$-Nuclei Collisions}
\vskip 0.5 cm
{\bf  T.Grigalashvili, L.Chikovani, E.Ioramashvili, A.Javrishvili,
     L.Khizanishvili, E.Mailian, M.Nickoladze, L.Shalamberidze}
\vskip 0.5cm
{\it Institute of Physics, Georgian Academy of Sciences,\\
 Tamarashvili st.6, Tbilisi, 380077 \\
 Georgia\\
  e-mail: eior@physics.iberiapac.ge\\
  e-mail: etheri@iph.hepi.edu.ge}
\end{center}
%\vskip 1cm
\begin{abstract}

The interactions of $\bar p$ with $D(2), Li(7), C(12), S(32), Cu(64)$ and
$Pb(207)$ nuclei at $40~GeV/c$ were studied by RISC-streamer chamber
spectrometer.  The yields of $K^0$ mesons and $\Lambda$ and $\bar\Lambda$
hyperons as functions of the target nucleus mass numbers are investigated.
The experimental results are compared with  model predictions using
$FRITIOF-7.02$ program package.
\end{abstract}

\section{Introduction}
\setcounter{equation}{0}
\par
 Investigation of strange
hadron formation at $\bar p p$ and $\bar p A$ interactions is of a great
interest by several reasons. On the one hand, as in any other hadron-hadron
collisions, the particles with nonzero strangeness can be formed only as a
result of sea $ s\bar s$ quarks production and of their subsequent
hadronisation under condition of open strangeness. The study of such
conditions gives the information on peculiarities of $s\bar s$-pair formation
and of their hadronisation mechanism.  \par On the other hand, availability
of various data on strange particle formation permits to carry out
comparative analysis of nonzero strangeness hadrons formation peculiarities
at $32~GeV/c$ $pp$ and $\bar p p$ interactions $[1-3]$.  This allows to
investigate the excess mechanism of strange mesons and barions formation at
$\bar p p$ collisions in comparison with $pp$ interactions.
The comparison of neutral strange particle formation gives evidence
concerning the statement about $pp$ and nonannihilated $\bar p p$
interactions equivalence [4].

The main contribution to the formation of new heavy quarks at
($\sqrt{s}<20~GeV$) intermediate energies  should give the light
valence quarks and antiquarks annihilation subprocesses:
$$
q_v \bar q_v \rightarrow g \rightarrow
s\bar s, c\bar c,...
$$

The large contribution of valence quarks annihilation to hard
subprocesses is established quite reliably. For example, the yield
of $J/\Psi$ particles at $40~GeV/c$ $\bar p p$ interactions exceeds
the yield at $pp$ interactions $\sim$ 12 times [5,6].

It was shown [3,7] that at $32~GeV/c$ the main excess in the
strange particle formation cross section at $\bar p p$ interactions is
connected with $\bar p p$ annihilation process.

The mesons multiplicity increases with energy in  $\bar p p$
annihilation processes, as at other hadron interactions [8],
and the valence quarks and antiquarks of initial nucleons can be a
part of different secondary mesons. For production of new $s\bar s$
pairs at $\bar p p$ annihilation can be spent a significant part of
total energy as compared to $pp$ interactions. This is caused by the
fact that at $pp$ interactions part of the total energy ($\sim$ 50 $\%$)
is consumed by leading quarks-spectators which, according to existing
knowledge, have a week influence on the process of secondary hadrons
formation [9].

But at $\bar p p$ annihilation all quarks and antiquarks of initial
system can take part in the process of new $s\bar s$ pair formation.
In quark-parton models, based on $QCD$ principles, such mechanism of
particle production at $\bar p p$ interactions is realized through
annihilation or exchange of more than one $q \bar q$ pairs. In quark
models based on the dual topological unitarization scheme $\bar p p$
annihilation is described by "three-chain" diagram [10] in which all
three $q \bar q$ pairs take part in creation of new $s \bar s$ pairs.

Present article deals with the experimental study of the strange
$K^0$ and $\Lambda$ and $\bar\Lambda$ particles generation process at
$40~GeV/c~\bar p A~(D, Li, C, S, Cu, Pb)$ interactions. The experimental
material was obtained on Relativistic Ionization Streamer Chamber
$(RISC)$ in magnetic field. Our RISC setup ensured  $4\pi$ angular
coverage.

The experimental results will be compared with model predictions
using $FRITIOF-7.02$ program package [11].

\section{Experimental apparatus}

The main part of $RISC$ spectrometer is a large  three-gap streamer
chamber $(4.7 \times 0.9\times 0.8)~m^3$ [12] placed inside a
magnetic field of about $1.5~T$.  The $HV$-pulses of $\pm$ 400 kV and
$\sim$ $20~ns$ duration were produced and shaped by bipolar Marx
generator and Blumlein line [13]. The chamber was filled  with helium -
neon gas mixture $(50\% Ne~ 50\% He)$ at atmospheric pressure. The
memory time of $(1-2)~\mu s$ was achieved by a slight admixture of
$SF_6$. The sensitive volume of the chamber was viewed by $8$ objectives
, each equipped with  two-stage image intensifier [14]. In addition, the
triggered events could be directly controlled by means of television
monitor.

The experiment was performed at Serpukhov proton synchrotron, using
an unseparated beam of negatively charged particles with momentum of
$40~GeV/c$. The beam was composed of $\pi^-$ and $K^-$ and $ \bar p $ in
$100:1.8:0.3$ ratio and had a momentum spread of $\Delta P/P\sim 1.5\%$.

The nuclear targets were placed in the visible volume of the chamber
along the beam line with the spacing of $30~cm$. The total thickness of
all targets corresponded approximately to $12\%$ of the proton
absorption length.  The elliptic target disks were mounted inside
cylindrical mylar boxes.

A telescope consisting of two scintillation
counters located behind the magnet of spectrometre served as a detector
designed to single out the events of nonelastic interactions of beam
particles in target. Signal of simultaneous starting  operation of
both counters was switched on for anticoincidence with the signal of
incident particles, forming a "trigger of interaction" of corresponding
particle in the target of spectrometer. On trigger signal the starting of
streamer chamber took place. The detector excluded the events with
negatively charged secondary particles for which the square
of transferred  4 momentum equals in average to
$t \leq 0.05 GeV/c^2 $. At the same time a great part of the
events of elastic scattering was excluded while the losses
of non-elastic events were equal to 3 $\%$ according to the
evaluations on the basis of the data of [15].  A more detailed
description  of the detector can be found in [16].

\section{Data analysis}

About $18 000$ frames of photographic film were scanned, and $7 489$
inelastic $\bar {p}$ interactions with deuterium, lithium, carbon,
sulphur, copper and lead nuclei were found. At the scanning stage
all secondary two-prong stars $V^0$ (further refereed to as vees)
were selected. They possibly are originated from the decays  of neutral
strange particles emitted from the primary interaction vertex. A double
scan of the film ensured  $\sim$ 99 $\%$ efficiency in detecting  the
vees.

All tracks  associated with the primary  vertex, as well as the vertex
and the  tracks of a vee, were first measured and then geometrically
reconstructed in the chamber volume. Events were retained for further
analysis if:

(a). the momentum error of the secondaries and of the $V^0$- decay tracks is
less then 10$\%$ and the residual (in space)  does not exceed 1800 $\mu$m;

(b). the primary vertex is reconstructed within the target;

(c). the $V^0$'s decay inside the fiducial volume. The cut imposes a target
dependent minimum $V^0$- path length and a maximum downstream length of 1m.
The radius of the fiducial volume  was chosen to ensure a
minimum track length of 15 cm.

(d). the space point reconstruction resolution in horizontal and vertical
planes  are  $\Delta$X$\leq$0.1 cm, $\Delta$Y$\leq$0.1 cm,
$\Delta$Z$\leq$0.4 cm.

If the angle between the momentum of the parent neutral particle and the
directions of two tracks of a vee was more then $ 6^0$, i.e.  noncoplanary,
this vee was rejected.  (The mean uncertainty in the noncoplanarity angle is
$\sim 0.7^0$). Vees rejected according to this criterion are mostly due to
two-prong inelastic interaction or three-body decays of neutral particles.

A vee passing the above coplanarity selection was then kinematically
fitted to each of four hypotheses ($K^0_S, \Lambda, \bar \Lambda,
\gamma$) by the method of least squares for three degrees of freedom
$(3DF-fit)$. To resolve the $V^0$/$\gamma$- ambiguity, photon conversions
were rejected by demanding M($e^+e^-$) $>$ 30 MeV/$c^2$. This cut is
sufficient to reject almost all $\gamma$'s and does not remove $V^0$'s. To
provide more reliable kinematical identification we retained only vees that
satisfied the following criteria:

(a). For each track of a vee, the root-mean-square deviation of measured
points from the fitted trajectory should not exceed $1.5~mm$ in the
chamber volume.

(b). For each track of a vee, the relative momentum error
$\Delta P/P \sim$ should be within $10\%$.

A vee was considered as an unambiguously identified if only one of
the four hypotheses had a $\chi^2$ value below $12$ (see Fig.1). The mean
values of $\chi^2$ for identified particles proved to be \begin{center}
 $<\chi^2>_{K_S^0} = 2.8 \pm$ 0.2,  $<\chi^2>_\Lambda = 3.0 \pm$
 0.2,\\
       $<\chi^2>_{\bar \Lambda} = 3.3 \pm$ 0.3.\\
\end{center}

A vee was considered as an ambiguously identified if the condition
$\chi^2 \leq $ $12$ was satisfied for two of the hypotheses. The
ambiguity between $\Lambda$ and $K_S^0$ was resolved in favor of
$K_S^0$, since the transverse momentum of the negatively charged
particle from the vee exceeded $105~MeV/c$ within the error. Observed
ionization along both tracks of a vee was also used in identifying
the parent that decayed. In total, nearly $5 \%$ of all detected neutral
strange particles could not be unambiguously identified.
These $V^o$ particles, determined ambiguously, were taken
into account in calculations of output, but they did not
participate in distribution of other kinematic characteristics.

Those neutral strange particles are indeed correctly identified that are
demonstrated in Fig.$(2\div 4)$. The distributions of $cos(\Theta^*)$ for
 identified strange particles are shown in Fig.2, where $\Theta^*$
 is the angle between the directions of the $V^o$ and one of the decay
products calculated in the $V^o$ rest frame for identified $V^o$-s.  The
 masses of $K^0$ and $\Lambda$ and $\bar \Lambda$ candidates are plotted in
 Fig.$3$.  The mean values are $<M_{K_S^0}>=(0.490 \pm 0.003)GeV$,
 $<M_\Lambda> =(1.115 \pm 0.001)GeV$ and $<M_{\bar \Lambda}> =(1.116
 \pm0.001)GeV$. The correlation between Podolansky-Armenteros parameter
 $\alpha $ and  the negative decay product transverse momentum with respect
 to the vee direction is illustrated in Fig.$4$.  The bands populated by
 neutral strange particles of each type are clearly seen.

To reconstruct the numbers of genuinely produced $K^0$ mesons,
$\Lambda$ hyperons and $\bar \Lambda$ hyperons, we introduced the
required correction factors.

$(1)$. The loss of events due to the limited dimensions of the chamber
and the loss of particles in the target were taken into account by means
of a geometric correction factor $<W_1>$, which depends on the
target type since the mean multiplicity of emitted particles grows
with increasing mass number of the target; as a result, the efficiency
of determination of  the vee vertex location deteriorates.
\vskip 0.5cm

\begin{center}
$W_1=1/[exp(L_1/L_0)-exp(L_n/L_0)]$
\end{center}

\vskip 0.5cm

\noindent $L_1$ is the radius of space in the vicinity of the target,
where the observation because of $V^o$ peak is impossible of
a large number of charged secondary particles. \\
$L_n$ is the potential path.\\
$L_0$ is the length of free path of vee.

$(2)$. Unobserved decay channels $K_S^0 \to \pi ^0  \pi ^0$,
$\Lambda \to n + \pi ^0$ and $\bar \Lambda\to \bar
n + \pi ^0$, as well as the emission of long-lived $K_L^0$ mesons,
were taken into account by means  of  $W_2(K^0)= 2.92$
$\pm 0.06$ and $W_2(\Lambda)$ = $1.56 \pm 0.01$ factors.

$(3)$. The correction for the efficiency of scanning was included
through factor $W_3 = 1.01$.

The overall correction factor $W$ represents the product of the
above mentioned three factors: $W =~<W_1>W_2W_3$. The values of the
geometric correction $<W_1>_{K_S^0}$, $<W_1>_\Lambda$,
$<W_1>_{\bar\Lambda}$ are presented in Table $1$.

\section{Experimental results}

The A-dependencies of the yields of neutral strange particles
$(K^0,\Lambda,\bar\Lambda)$ were investigated  in the following
processes:
\begin{equation}
    \bar{p} A \to K^0(K^0Y,K\bar K) + X
\end{equation}
\begin{equation}
\bar{p} A \to \Lambda(\Sigma^0) + X
\end{equation}
\begin{equation}
\bar{p} A \to \bar \Lambda (\bar \Sigma^0) + X
\end{equation}
where $Y$ is the hyperons ($ \Lambda,\bar\Lambda,\Sigma^0,\Sigma^\pm $).

The experimental statistics, used in present article, is shown in
Table $1$. It includes the number of inelastic $N_{int}$ interactions,
among which $V^0$ events were found and the number of unambiguously
identified neutral strange $(K_S^0, \Lambda, \bar \Lambda)$ particles.

After making corrections for losses of $ { V^0}$  events, connected with
restrictions of mean path of particles in chamber and with nonobserved
mode of decay of $ {K^0} $ mesons and $ {\Lambda}$ and $ {\bar\Lambda}$
hyperons, the total number of neutral strange particles for all studied
nuclei were calculated . Thus, $ N _{\Lambda}$ includes also ${\Lambda}$
hyperons, which have arisen as a result of $\Sigma ^0$ decay.

On the basis of these data the average $ <N_{K^0}>$, $<N_{\Lambda}>$
$ <N_{\bar\Lambda}>$ magnitudes presented in Table 2 were calculated for
one inelastic interaction. The inclusive multiplicity $ <N_{K^0}>$ takes
into account contributions both from the production of kaon pairs
($K^0\bar K^0,K^0K^-$ and $\bar K^0K^+$) and from associated kaon-hyperon
production ($K^0\Lambda, K^0\Sigma^0,K^0\Sigma^-,\bar K^0\Sigma^+$ and $\bar
K^0\bar\Lambda$). For their comparison the values calculated by Monte-Carlo
method by program $FRITIOF-7.02$ are presented for $\bar p A$ interactions.
The table gives also the average magnitudes of neutral strange particle
yields, obtained for $\bar {p} p$ interactions at $32~GeV/c$ momentum [3] and
at $100~GeV/c$ momentum [17].  As it is shown from the Table the
experimental results (for $K^0$ and $\Lambda$) slightly exceed the results
of model. At increasing the mass number of target nucleus from deuterium to
lead the average magnitudes of particle yields increase:  $<N_{K^0}> - 1.8$
times, and $ < N_{\Lambda}> - 2.5$ times; but  $ <N_{\bar \Lambda}>$
remains constant within the error. The average magnitudes of $\Lambda$ and
$\bar\Lambda$ yields  are equal both  on hydrogen and on deuterium [3,17],
as far as the C-symmetry at $\bar {p} p$ interactions assumes the equality
of probabilities of $\Lambda$ and $\bar\Lambda$ hyperons generation in
events with identical multiplicity.

The $A$-dependencies of yields of ${K^0}$ mesons and $\Lambda$ and
$\bar\Lambda$ hyperons generated by reactions $(1,2,3)$ are shown
in Fig.$5(a,b,c)$.  The solid lines are the results of fitting the
experimental data and dashed lines - the similar events calculated by
$FRITIOF-7.02$ program with power function $y = aA^\alpha $. The
parameters ($a$, $\alpha$) and  $\chi^2$ are presented in Table $3$.

The behavior of yields of neutral strange particles with the increase of
mass number is shown more clearly on the plot of dependence of relative
$R^{A}_{V^o}$ yield (ratio of $N_{K^0 }, N_{ \Lambda } $ and $N_{
\bar\Lambda}$ for nucleus and corresponding values for hydrogen) and the
ratio $R_{V^0}=\frac{<N_{K^0}>}{<N_{\Lambda}>} $ and
$\frac{<N_{\bar\Lambda  }>}{< N_ {\Lambda }>}$ on
$<\nu_{ eff}>$ $($fig.6 a,b$)$, where $<\nu_{eff}>$ is the effective
number of interactions in the frameworks of Glauber-Gribov model [18,19],
and is determined as
\begin{center}
  $<\nu_{eff}>$ = $<\nu>/<\nu>^a$
\end{center}

and  $<\nu>^a  = A\sigma^a_{\bar p p }/\sigma^a_{ \bar pA}$ - is the
correction for annihilation  channels, where $ \sigma^a_{\bar p p}$ =
$5.6~mb$, is taken as the difference between sections of $\bar {p} p$
and $pp$ interactions at $40~GeV/c$ [20].

The ratio of cross sections for annihilation and inelastic processes
$\sigma^a_ {\bar p A} / \sigma^{inel} _ {\bar p p}$ calculated in the
framework of Glauber-Gribov model [21] is given in Table $4$.

All experimental results in Fig.$4 (a,b)$ are fitted
by linear functions $y = a + b<\nu_{ eff}>$, the parameters of
which are given in Tables $5$.

Thus, within the experimental error
$R^A_{\Lambda}=\frac{<N_\Lambda> \bar p A}{<N_\Lambda> \bar p
p}\approx<\nu_{eff}>$, and $R^A_{\bar \Lambda}=
\frac{<N_{\bar \Lambda}>\bar p A}{<N_{\bar \Lambda}>\bar p p}\approx 1$.
More rapid fall of $R_{K^0,\Lambda}= \frac{<N_{K^0}>}{<N_{\Lambda}>}$ is
observed as compared to $R_{\bar \Lambda,\Lambda}=
\frac{<N_{\bar\Lambda}>}{<N_{\Lambda}>}$.

\section{Discussion of results}

The obtained results were compared  with those on $FRITIOF-7.02$ based
on quark-gluon-string  model [11]. The universality of strings
hadronisation is assumed in this model. This means the identity of
multiple processes, taking place at inelastic $pp$ and nonannihilation
$\bar{p}p$ interactions, since the processes of annihilation and
diffraction are absent in Lund model.

The model describes a wide sphere of phenomena from formation of
charmed and beauty particles and effects of the fine gluons in hard
processes to production of particles in soft interactions of hadrons
with nucleons and nuclei [22-24].

In  $FRITIOF-7.02$ (version [25,26]) based on Lund model for
soft hadron-hadron collisions the longitudinally-excited state is
formed, i.e. the string between valence quarks (antiquarks) and diquarks
(antidiquarks) of each hadron is stretched.  The string is fragmented
into hadrons by breakage.  In the case of hadron-nuclear interaction
projectile hadron interacts more than once and obtained objects in
excited states collide with nucleons before they passing through
nucleus. Since the hadron fragmentation time is more than the
internucleonic distance the string stretched by projectile particle
does not have a time for fragmentation and continues to interact with
nucleons of nucleus. As a result we have some number of strings
fragmented into hadrons outside the nuclei.

The comparison of present experimental results with $FRITIOF-7.02$
calculations shows a small difference in the neutral strange
particle yields. This can be explained by the fact that $\bar{p}p$
annihilation is not taken into account in the model.

In experiments on heavy nuclei the value of $N_{K^0}$ yield should be
even more and the value of $N_\Lambda$ yield should be approximately the
same as in calculations since the probability of annihilation increases
with increasing the atomic number [21,27].

Thus, the  experimental results can be interpreted as follows: some
differences in yields of $K^0$ mesons between the model and the
experimental data is the consequence of the existence of annihilation
channel at $\bar{p}A$ interactions.  At the same time, a part of $K^0$
mesons (produced basically in annihilation channel) has time for
secondary interactions in nucleus. This leads to the increase of
$\Lambda$ hyperon yield, as the threshold energy of $\Lambda $
production is $\sim 2~GeV/c$ [28].

The experiment shows that $\bar\Lambda$ hyperon yield, as was mentioned
above, is practically independent of the atomic number, though the model
predicts their small increase. Within the error their yield is
equal to that of $\Lambda$ hyperons for hydrogen and deuterium, and
coincides with model prediction. This can be explained as follows:

The probability  of annihilation processes increases with increasing
the atomic number. But, on the one hand, this does not give the
contribution to production of  $\bar\Lambda$ hyperons. On the
other hand, $K^0$ mesons, produced in annihilation channel, can not
create $\bar\Lambda$ hyperons because of large threshold energy
($\sim 7~GeV$ [28]).

The problems connected with other characteristics of $\bar{p}A$
interactions with production of neutral $K^0$ and $\Lambda$ and
$\bar\Lambda$ particles will be considered in further publications.
Namely, the associated multiplicities of nucleons and negative mesons
to the neutral strange particles; the inclusive characteristics of
neutral strange particles and their dependencies on mass number of target
nuclei will be investigated.

\section{Conclusion}

In the present article we have investigated the production of neutral
strange particles in the processes of $\bar{p}A$ collisions at incident
$40~GeV/c$ momentum  in wide range of nuclear targets $(D, Li, C, S, Cu,
Pb)$. The yields of $K^0$ mesons and  $\Lambda$ and $\bar\Lambda$
hyperons have been defined in dependence on mass number of nuclei in
reactions  $(1,2,3)$.

The experimental results have been compared with theoretical predictions
based on $FRITIOF-7.02$ code. Following results should be mentioned:

1. The yield of  $K^0$ mesons in process $(1)$ grows as $\sim
(0.20 \pm 0.01) A^{(0.141\pm 0.007)}$.

2. The yield of $\Lambda$ hyperons in process $(2)$ grows as $\sim
(0.054 \pm 0.005)A^{(0.22 \pm 0.01)}$ .

3. The yield of $\bar\Lambda$ hyperons in process $(3)$
remains   constant within experimental error $\sim (0.07 \pm 0.01
) A^{(-0.02 \pm 0.02)}$. Thus the relative yields $R^A_{\Lambda}=
\frac{<N_\Lambda> \bar p A}{<N_\Lambda> \bar p p}\approx<\nu_{eff}>$
and $R^A_{\bar\Lambda}=
\frac{<N_{\bar\Lambda}> \bar p A}{<N_{\bar\Lambda}>\bar p p}\approx 1$,
where $<\nu_{eff}>$ is the effective number of interactions in nuclear.

4. Experimental values of $K^0$ and $\Lambda$
particles yield slightly exceed the similar value predicted by
$FRITIOF-7.02$ code, but their $A$ dependencies
coincide  within the error.\\
$\sim (0.179 \pm 0.004) A^{(0.155 \pm 0.003)}$ for $K^0$ mesons,
\\ $\sim (0.054 \pm 0.003) A^{(0.202 \pm 0.007)}$ for $\Lambda$
hyperons. \\
There is difference only in yields of $\bar\Lambda$
hyperons, where by $FRITIOF-7.02$ a growth of $\sim (0.060 \pm 0.002) A^
 {(0.074 \pm 0.004)}$  is predicted.

\section{Acknowledgement}

We thank  our  colleagues in  $"RISC"$ Collaboration for their
contribution to all stages of the  experiment. We are indebted to
N.N.Roinishvili and J.Manjavidze for enlightening comments and
stimulating discussions.

This work was supported in part by the  Academy of Sciences of Republic
Georgia (Grant no. 2.11).

\newpage

{\bf{Table Captions}}
\par
{\bf {Table 1.}} Sum of experimental statistics and geometric
correction factors.
\par
{\bf {Table 2.}} Observed and predicted mean multiplicities of inclusive
$K^0$ mesons  and  $\Lambda$ and $\bar \Lambda $ hyperons.
\par
{\bf {Table 3.}} The parameters $(a,\alpha)$ and $\chi^2$ of
$y = aA^\alpha$ function for yields of neutral strange particles for
experimental results and model predictions.
\par
{\bf  {Table 4.}} The values of ratio of the annihilation and the
inelastic cross sections $\sigma^a_{\bar p A}/\sigma^{inel}_{\bar p p}$
and corresponding $<\nu_{eff}>$ for nuclei.
\par
{\bf { Table 5.}} The parameters $($a,b$)$ and $\chi^2$ of $y = a +
b<\nu_{eff}>$ function for $R^A_{V^0}$ and $R_{V^0}$ ratios for
experimental  results.

\newpage

{\bf{Figure Captions}}
\par
{\bf{Fig.$1$.}} Distribution of $P(\chi^2)$  for unambiguously
identified particles.
\par
{\bf{Fig.$2$.}} Distribution of $cos(\Theta^*)$  for unambiguously
identified particles.
\par
{\bf{Fig.$3$.}} Distribution of invariant mass  for
identified particles (for $K^0_S \to M(\pi^+ \pi^-)$, for $\Lambda\to M(p
\pi^-)$ and for $\bar \Lambda\to M(\bar p \pi^+)$).
\par {\bf{Fig.$4$.}}
Scatter plot of transverse momentum $(p_t)$ of negative particles as a
result of decays of $K^0_S$ and $\Lambda$ and $\bar \Lambda$ versus
Podolansky-Armenteros parameter $(\alpha)$.
 \par {\bf{Fig.$5(a,b,c)$.}}
Mean multiplicities of (a) inclusive $K^0$ mesons, (b) $\Lambda$ and (c)
$\bar \Lambda$ hyperons as function of the mass number of the target
nucleus: $(o)$ experimental data and $(*)$ FRITIOF-7.02 predictions.
 \par
{\bf{Fig.$6(a,b)$.}} $R^A_{V^0}$ and $R_{V^0}$ as functions of
$<\nu_{eff}>$ (see in the text).

\newpage
\begin{center}
{\bf{Table 1}}. Sum of experimental statistics and geometric
correction factors
\end{center}
\begin{sloppypar}
\begin{center}
\begin{tabular}{||c||c|c|c|c|c|c|c||}
\hline \hline
 A &$N_{int}$ & $N_{K^0_S}$ & $N_{\Lambda}$ & $N_{\bar \Lambda}$ &
 $<W_1>_{K^0_S}$ & $<W_1>_{\Lambda}$ & $<W_1>_{\bar \Lambda}$\\
%      &    &   &    &    &    &     &
\hline \hline
 D& 1741 & 35 & 23 & 27 & 3.6$\pm$.4 &3.5$\pm$.5 &2.9$\pm$.3\\
%    &    &   &    &    &    &     &
\hline
 Li & 1149 & 36 & 22 & 20 &2.4$\pm$.3 &2.2$\pm$.4 &2.0$\pm$.3\\
%    &    &   &    &    &    &     &
\hline
 C &1053 & 30 & 16 & 16 &2.8$\pm$.4 &3.2$\pm$.4 &2.0$\pm$.2\\
%     &    &   &    &    &    &     &
\hline
 S & 1197 & 47 & 36 & 24 &2.6$\pm$.2 &2.4$\pm$.2 &2.3$\pm$.2\\
%    &    &   &    &    &    &     &
\hline
 Cu & 1346 & 42 & 34 & 14 &3.6$\pm$.5 &3.4$\pm$.4 &2.8$\pm$.4\\
%    &    &   &    &    &    &     &
\hline
 Pb & 1003 & 29 & 28 & 17 &3.8$\pm$.6 &3.6$\pm$.4 &2.1$\pm$.4\\
%    &    &   &    &    &    &     &
\hline \hline
\end{tabular}
\end{center}
\end{sloppypar}
\vspace{0.5cm}
%\newpage

\begin{center}
{\bf{Table 2.}} Observed and predicted mean multiplicities of inclusive
$K^0$ mesons and $\Lambda$ and $\bar \Lambda$ hyperons.
\end{center}
\begin{sloppypar}
\begin{center}
\begin{tabular}{||c||c|c|c||}
\hline \hline
 Sample     & $<N_{K^0}>$ & $<N_\Lambda>$ & $<N_{\bar \Lambda}>$ \\
\hline \hline
 $\bar p p[3]$ 32 GeV/c  & 0.21$\pm$.02 & 0.05$\pm$.01 & 0.06$\pm$.01 \\
 $\bar p p[17]$ 100 GeV/c & 0.30$\pm$.02 & 0.07$\pm$.01 & 0.07$\pm$.01 \\
  Prediction & 0.1902 & 0.0633 & 0.0654 \\
\hline
 $\bar p D$   & 0.24$\pm$.02 & 0.07$\pm$.01 & 0.07$\pm$.01 \\
  Prediction & 0.1997 & 0.0585 & 0.0628 \\
\hline
 $\bar p Li$   & 0.25$\pm$.02 & 0.07$\pm$.01 & 0.06$\pm$.01 \\
  Prediction & -- & -- & -- \\
\hline
 $\bar p C$   & 0.27$\pm$.02 & 0.08$\pm$.02 & 0.06$\pm$.02 \\
  Prediction & 0.2567 & 0.0941 & 0.0741 \\
\hline
 $\bar p S$   & 0.35$\pm$.02 & 0.13$\pm$.01 & 0.07$\pm$.01 \\
  Prediction & 0.3148 & 0.1118 & 0.0752 \\
\hline
 $\bar p Cu$   & 0.38$\pm$.02 & 0.15$\pm$.02 & 0.05$\pm$.02 \\
  Prediction & 0.3385 & 0.1240 & 0.0791 \\
\hline
 $\bar p Pb$   & 0.42$\pm$.03 & 0.17$\pm$.02 & 0.06$\pm$.02 \\
  Prediction & 0.4069 & 0.1503 & 0.0911 \\
\hline \hline
\end{tabular}
\end{center}
\end{sloppypar}
\vspace{0.5cm}
\newpage

\begin{center}
{\bf{Table 3.}} The parameters $(a,\alpha)$ and $\chi^2$ of $y = aA^\alpha$
function  for yields of neutral strange particles for experimental
results and model predictions\\
\end{center}
\begin{center}
\begin{tabular}{||c||c|c|c||}
\hline \hline
\Large Process & \Large a &  \Large $\alpha$ & $\chi^2$/NDF \\
\hline \hline
$\bar p A\to K^0(K^0Y)$ + X  &&&\\
Experiment &  0.20$\pm$.01 & 0.141$\pm$.007 & 1.3 \\
Prediction &  0.179$\pm$0.004  & 0.155$\pm$0.003 &  \\
\hline
$\bar p$A$\to\Lambda(\Sigma^0)$ + X &&& \\
Experiment &  0.054$\pm$.005 & 0.23$\pm$.01 & 0.7\\
Prediction &  0.054$\pm$0.003 & 0.202$\pm$0.007& \\
\hline
$\bar p$A $\to\bar \Lambda (\bar \Sigma^0)$ + X &&& \\
Experiment &0.07$\pm$.01 & $-$0.02$\pm$.02 & 0.4  \\
Prediction & 0.060$\pm$0.002 & 0.074$\pm$0.004 & \\
\hline \hline
\end{tabular}
\end{center}
\vspace{0.5cm}

\begin{center}
{\bf{Table 4.}}The velues of ratio of the annihilation and the inelastic
cross sections $\sigma^a_ { \bar p A } / \sigma^ { inel } _ { \bar p p}$
and corresponding $<\nu_{eff}>$ for nuclei\\
\end{center}
\begin{sloppypar}
\begin{center}
\begin{tabular}{||c||c|c|c|c|c|c|c||}
\hline \hline
 A & H & D & Li & C & S & Cu & Pb \\
\hline \hline
 $\sigma{^a_{\bar p A}}/\sigma{^{inel}_{\bar p A}}$& 0.15 & 0.17 & 0.21
& 0.24 & 0.30 & 0.35 & 0.44 \\
\hline
$<\nu_{eff}>$ & 1.0 & 1.1 & 1.4 & 1.6 & 2.0 & 2.5 & 3.0 \\
\hline \hline
\end{tabular}
\end{center}
\end{sloppypar}
\newpage

\begin{center}
{\bf{Table 5.}} The parameters $($a,b$)$ and $\chi^2$ of $y = a +
b<\nu_{eff}>$ function for $R^A_{V^0}$ and  $R_{V^0}$ ratios for experimental
results\\ \end{center} \begin{sloppypar} \begin{center}
\begin{tabular}{||c||c|c|c||}
\hline \hline
 R &$ a$ & $b$ & $\chi^2$/NDF \\
\hline \hline
&&&\\
$R^A_{K^0}=\frac{<N_{K^0}>_{\bar p A}}{<N_{K^0}>_{\bar p p}}$ & 0.57
$\pm$0.06 & 0.53$\pm$0.02 & 0.2 \\ &&&\\ \hline &&&\\
$R^A_\Lambda=\frac{<N_\Lambda>_{\bar p A}}{<N_\Lambda>_{\bar p p}}$ &
-0.155$\pm$0.004 & 1.07$\pm$0.05 & 0.5 \\
&&&\\
\hline
&&&\\
$R^A_{\bar \Lambda}=\frac{<N_{\bar \Lambda}>_{\bar p A}}
{<N_{\bar\Lambda}>_{\bar p p}}$ & 1.2$\pm$0.2 &-0.06$\pm$0.03 & 0.5 \\
&&&\\
\hline \hline
&&&\\
$R_{K^0,\Lambda}=\frac{<N_{K^0}>}{<N_\Lambda>}$ & 4.1$\pm$0.8 &
-0.557$\pm$0.004 & 0.2 \\
&&&\\
\hline
&&&\\
$R_{\bar \Lambda,\Lambda}=\frac{<N_{\bar \Lambda}>}{<N_\Lambda>}$ &
1.4$\pm$0.2 & -0.36$\pm$0.02 & 0.3 \\
&&&\\
\hline \hline \end{tabular}
\end{center}
\end{sloppypar}

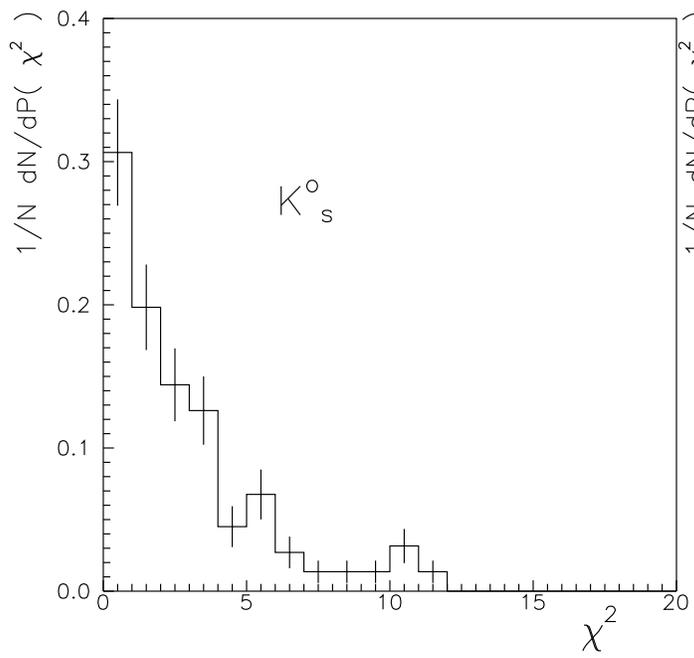
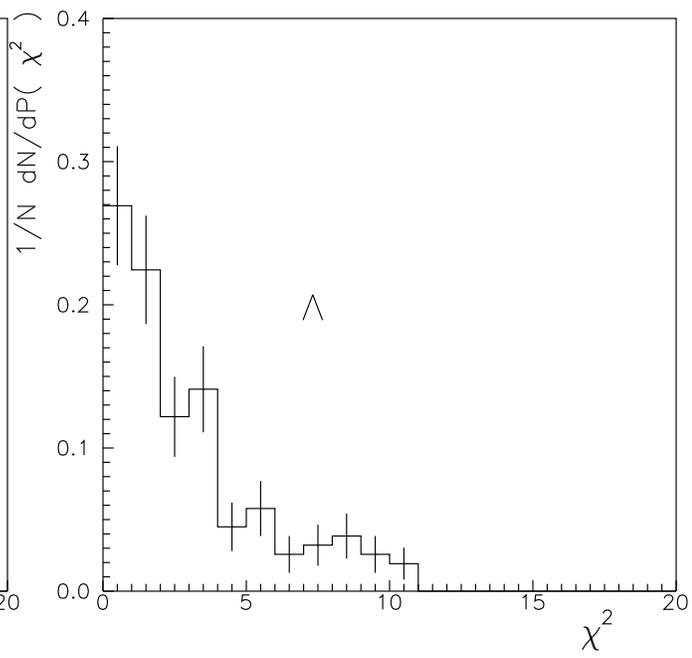
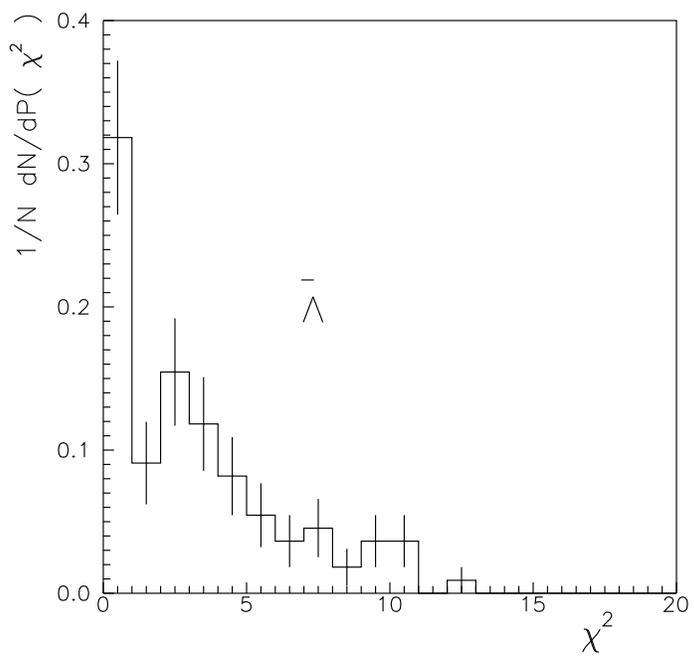
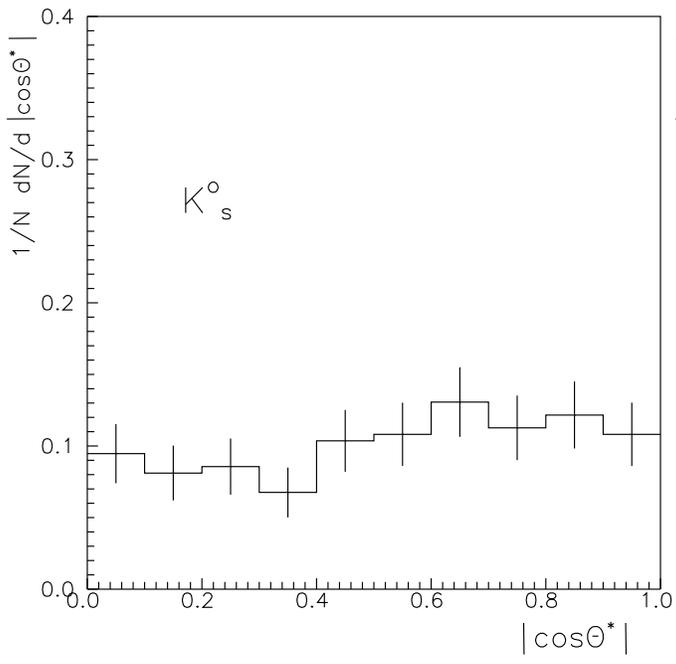
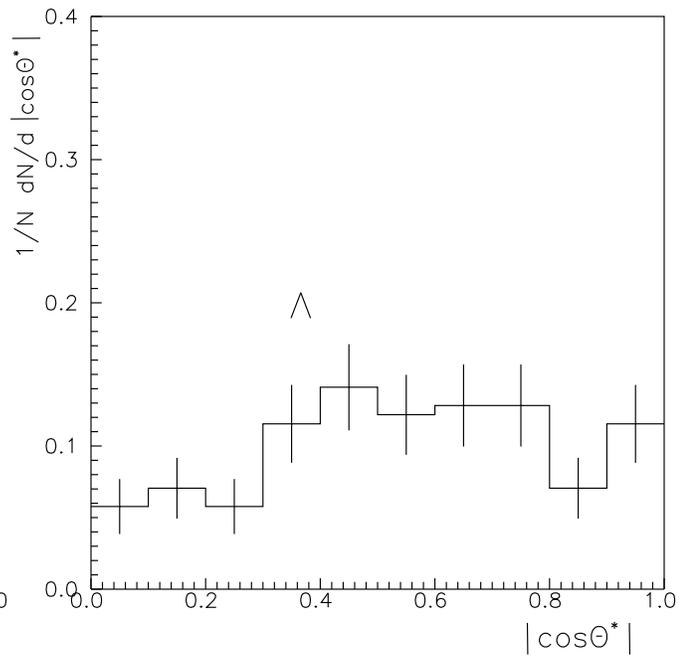
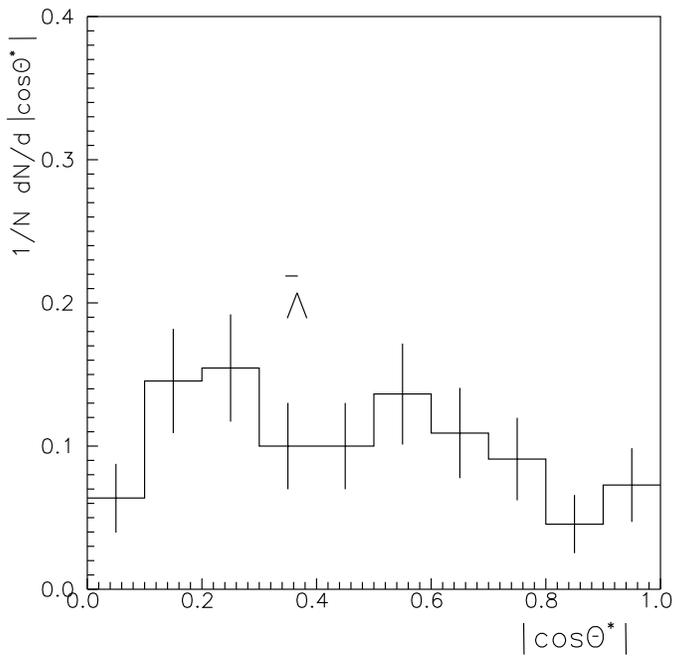
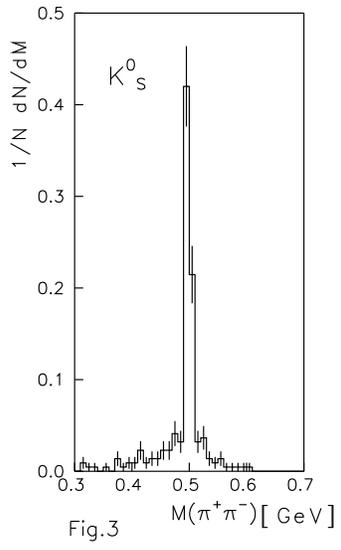
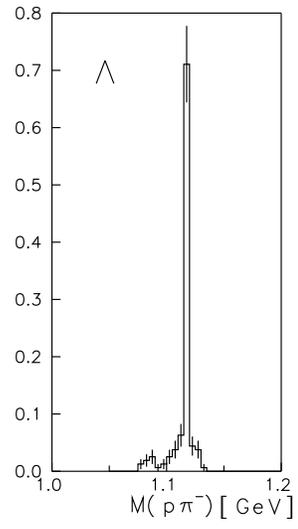
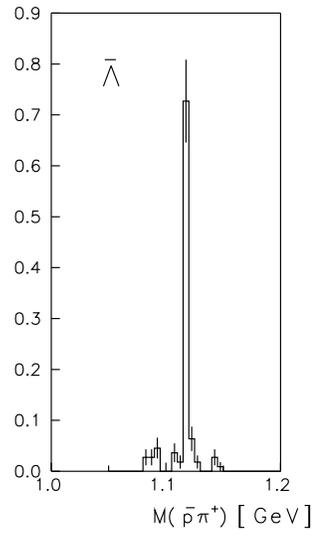
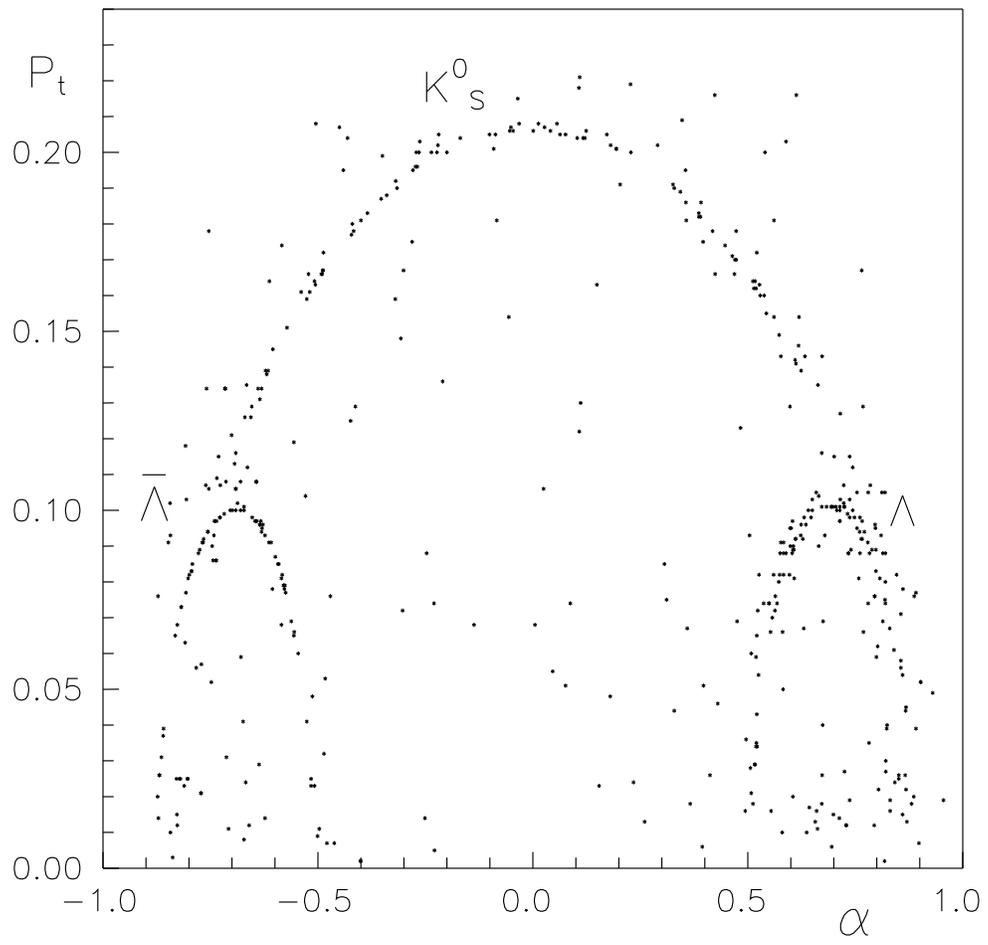
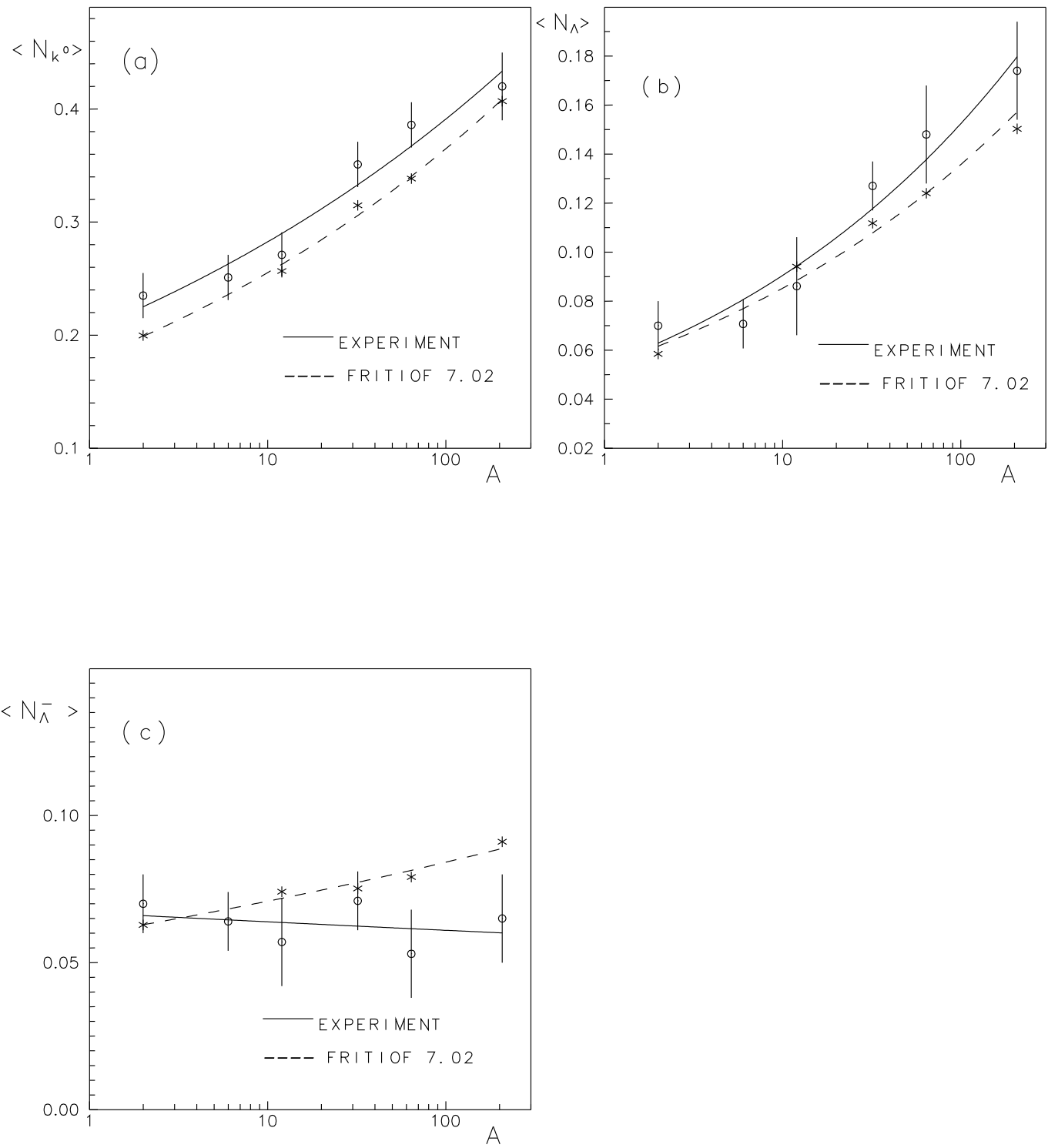
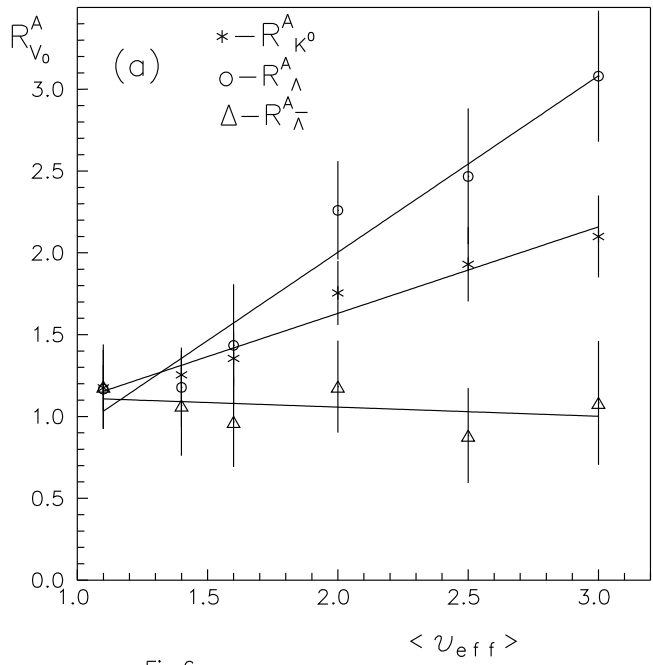
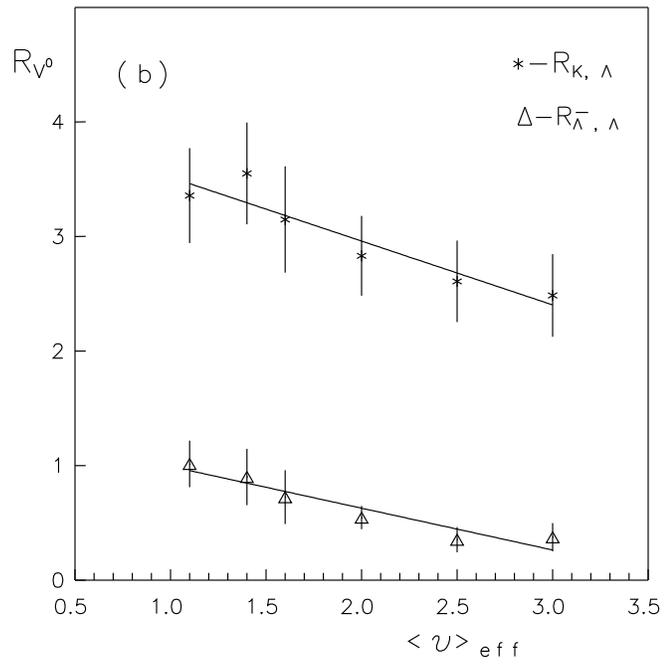
\end{document}